\documentstyle[aps,epsf,multicol,prl,epsf]{revtex}
\preprint{IASSNS-HEP-97/115}

\begin{document}
\title{An Order Parameter for the Mott-Hubbard Transition in One Dimension}
\draft
\author{S. P. Strong}
\address{School of Natural Sciences, Institute for Advanced Study,
Olden Lane, Princeton, NJ 08540}
\author{J.C. Talstra}
\address{James Franck Institute of the University of Chicago, 
5640 S. Ellis Ave., Chicago, IL 60637}
\date{October 9, 1997}
\maketitle
\begin{abstract}

We propose an order parameter for the one dimensional
Mott-Hubbard transition and provide numerical evidence
and general theoretical arguments for the
correctness of our proposal.  In addition,
we discuss some of the implications of this
picture of the one dimensional Mott-Hubbard transition and
speculate about possible higher dimensional analogs.

\end{abstract}


\begin{multicols}{2}

The one dimensional Hubbard model:
\begin{equation}
H_{\rm 1D ~Hubb}  =  t \sum_{ x,\sigma,\nu= \pm 1}
 c_{\sigma,x+\nu}^{\dagger} c_{\sigma,x} +
U \sum_{x} n_{\uparrow,x} n_{\downarrow,x}
\end{equation}
is known to exhibit
a $T=0$ quantum transition between metal and insulator
as a repulsive interaction in the model is turned on
\cite{liebwu} and an analogous transition
is believed to occur in higher dimensions
\cite{mott}.  These
``transitions'' are somewhat peculiar in that a
physical, local order parameter has not
been identified.  Instead, following
Kohn's suggestion \cite{kohn}, the ``order parameter''
for these transitions is generally 
taken to be the weight in the zero frequency delta 
function in the electrical conductivity at zero temperature.
In the metallic phase this weight is finite and
in the insulator it vanishes, so that the proposed
order parameter does in fact distinguish the
two phases qualitatively as it should.
However, the conductivity is clearly not locally defined,
so that it is meaningless to speak of, for example, a diverging
length scale in its correlation functions, and
it leads to a qualitative distinction between metal and
insulator only at zero temperature.  In
the limit of large spatial dimensions,
the Hubbard model undergoes a finite temperature
metal-insulator transition associated with a
crossing of the free energies of two distinct
phases \cite{georgesfootnote}; however,
there is no order parameter
distinguishing the phases in that case,
and the transition line vanishes at a critical point,
much like the liquid-gas transition of water.
This is not expected to be the case for the metal
insulator transition in one spatial dimension and
may also not be generic for finite dimensions. 
Therefore, it would be desirable to define some local
operator whose expectation value or correlations clearly distinguished
the metallic and insulating phases;
the purpose of this paper is to propose an operator
which fulfills these requirements for the one-dimensional
transition and to examine possible implications
for higher dimensions.

The operator we propose is an extension of the
operator introduced by us and P. W. Anderson
\cite{usprl} as exhibiting off-diagonal long range
order (ODLRO) in a wide variety of quantum spin chains.
There we noted that
the groundstate of a generic, gapless $XXZ$ model on $N+2$ 
sites has a finite overlap in the thermodynamic limit with
the state obtained by taking the groundstate of the
same model on $N$ sites and tacking a singlet pair of
spins onto the end. If we define an operator, $O^{\dagger}(i)$,
which adds a pair of sites with their spins in a singlet
configuration at site $i$ into spin model,
and its conjugate operator, $O(i)$, which
removes a pair of sites, $i$ and $i+1$, from 
the model if they are in a singlet configuration 
or else annihilates the wavefunction on which it
acts, then, as a consequence of the overlap mentioned above,
this operator has ODLRO in its correlation functions:
$\lim_{|i-j|\rightarrow\infty}\langle O^{\dagger}(i)
O(j) \rangle \neq 0$ \cite{likeFQHE}.

However, $\lim_{|i-j|\rightarrow\infty}\langle
O^{\dagger}(i) O(j) \rangle \neq C_1$;
rather, since the groundstates of generic $XXZ$ models
on $N$ sites and $N+2$ sites have momenta differing by $\pi$,
$\langle O^{\dagger}(i) O(j) \rangle \sim
(-1)^{i-j}$ and
our order parameter 
exhibits a broken $Z_2$ symmetry.  This $Z_2$  is unconnected
with the division of a Neel
ordered state on a bipartite lattice into sublattices:
there is no Neel order in one dimension even at zero 
temperature \cite{mermin}.
However, the theorems of Mermin and Wagner and
Coleman \cite{mermin} do allow the breaking of a discrete symmetry,
such as the $Z_2$ breaking indicated by our order parameter.

What, then, is the nature of this broken symmetry?
In \cite{usprl}, we utilized the connection between the
Heisenberg model and the large $U$ Hubbard model
to motivate the conjecture that the insertion
of a single spin of spin species $\sigma$
into a Heisenberg model at site $j$ was
equivalent in the Luttinger liquid description of that
models low energy physics to the action of the
operator $i^j \exp[\frac{i}{2} \Theta_{R,\sigma}(ja)] + 
(-i)^j \exp[\frac{i}{2} \Theta_{L,\sigma}(ja)]$.
This not only accounts for the ODLRO of the singlet insertion,
but explains the behavior of the family of operators in
which singlet pairs of spins are inserted into sites
which are not adjacent.  We have further investigated this
proposal using exact diagonalization and Haldane \cite{haldise}
and Shastry's \cite{shastryise}
solution of the inverse squared exchange model, to show that
spin insertion is essentially equivalent to spinon creation
\cite{usspinon} {\em or\/} the annihilation of a spinon
of the opposite spin species and the
conjecture is now firmly supported.   However, it
leads to a rather peculiar picture for the cause of the
broken symmetry.  The ``operator'' responsible in
Luttinger liquid description is $(-1)^j$, i.e.
a $c$ number \cite{otherspinorders}.  This suggests that the order is not
the order of a spin model at all, 
and we propose that
it should rather be thought of as the charge ``order'' which sets
in when a Mott-Hubbard transition occurs in one dimension
and the low energy effective theory is transformed from one
of interacting fermions to one containing only 
spin degrees of freedom.  In support of this,
we note that all of the spin models
for which we have demonstrated this order 
can be obtained as the low energy limits of fermion
models which have undergone such a transition.  

\begin{figure}
\narrowtext
\centerline{ \epsfxsize = 2.5in
\epsffile{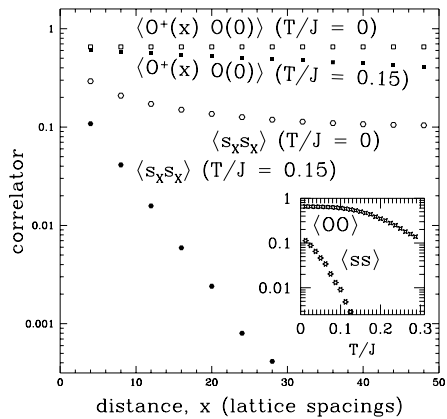}}
\caption{Zero and finite temperature results for the singlet insertion/deletion
correlator and the more conventional two point function
of $s_{\rm x}$.  Results are for a one dimensional $XY$ model of
100 sites with periodic boundary conditions (PBC). Inset: results
for correlators for a fixed separation of 25 sites as a function of temperature,
$T$.
}
\label{fig:xy}
\end{figure}

To check the proposal that the order in question
is really in the charge sector, we have carried out several numerical
and analytic investigations.  First, we have verified that
the ``order'' probed by the singlet insertion is not
that of the spin degrees of freedom by conducting a 
{\em finite temperature\/} study of the singlet insertion-singlet
deletion correlation function for the $XY$ model.  In that case,
the mapping of the model onto free fermions allows us to study
large systems sizes at finite temperatures to examine how
the spin and $Z_2$ orders are degraded.  Some results from
Monte Carlo are shown in Fig. \ref{fig:xy}.  
It is clear that the order is not  
degraded by finite temperature
in the same way as the more conventional spin correlations are.

\vspace*{-0.6cm}

Apparently, the identification of the singlet insertion
with a $c$ number in the spin model does not hold rigorously
and the singlet-correlation/singlet-deletion
correlator is affected by the finite temperature in the
spin model, lacking ODLRO at high and probably all finite temperature.
Nonetheless, the degradation is 
enormously less severe than
that occurring for other, typical spin correlation functions.
For example, at $T = .15 J$,
$\langle s^x s^x \rangle$ decays exponentially with
a correlation length of $\sim 4.0$ lattice spacings, while
$\langle O^{\dagger} O \rangle$ decays by well less than a
factor of $2$ over a separation of 50 sites! It is not
even clear from results at this temperature that the ODLRO
has been destroyed.  We attribute this difference in
behavior from conventional spin correlators as the 
result of the fact that 
the $XY$ model,  viewed as a model with an Mott-Hubbard 
ordered charge sector, has an infinitely large charge gap.
Thus the charge order, as probed by our singlet insertion,
is unaffected by finite temperature, which rapidly destroys all
of the usual spin order; the singlet insertion/deletion 
correlator is only indirectly connected with the charge sector
order and it appears that only if the spin degrees of 
freedom are effectively at zero temperature does it exhibit
ODLRO, so its correlator does decay at finite temperatures,
but extremely slowly.

Having shown that our order is almost not degraded for separations
and temperatures where spin order is totally lacking,
it remains to show that it is sensitive to charge
ordering of the Mott-Hubbard type.  First,  it is possible
to show that, for free, $SU(2)$, spin $\frac{1}{2}$
fermions at half filling, the operator which adds two
adjacent, occupied sites with spins in a singlet configuration
does not have ODLRO.
Rather, the correlation function decays algebraically
like $x^{-1}$ (or, for $SU(N)$ generalizations,
like  $x^{-\frac{N}{2}}$ for $N$ inserted
sites in an $SU(N)$ singlet).  We have shown this both numerically
and analytically \cite{usunpub}, and therefore know that free fermion systems,
which should not show Mott-Hubbard order, do not show 
ODLRO as probed by the singlet insertion operator.

\vspace*{-0.6cm}
\begin{figure}
\narrowtext
\centerline{ \epsfxsize = 2.5in
\epsffile{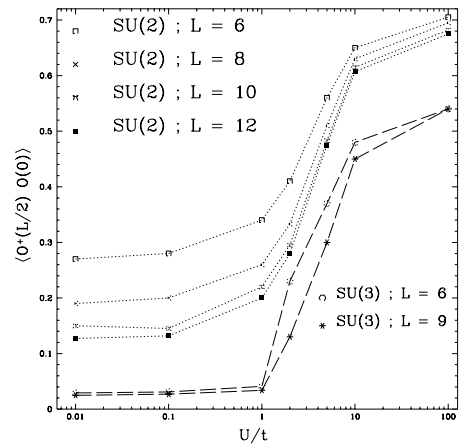}}
\caption{Singlet insertion/deletion correlators
for $SU(2)$ and $SU(3)$ Hubbard models of various
sizes (PBC).  Insertion and deletion were separated by
half the system size, $L$.}
\label{fig:su2su3}
\end{figure}

Conversely, if we consider the large $U$ limit of the
one dimensional Hubbard model at half filling,
then the singlet insertion operator clearly does
display long range order. This can be seen from the
Ogata-Shiba factorized wavefunction \cite{ogatashiba} 
which has a charge wavefunction given by that of spinless
fermions and a spin part given by the groundstate of the
Heisenberg model with the ``sites'' defined as the positions
of the spinless fermions.  At half filling, there is a fermion
on every site and the charge part is trivial, a property
unchanged by inserting or deleting singly occupied sites.
For this reason the singlet insertion correlations,
are identically those of the Heisenberg model and have ODLRO,
as we have previously established \cite{usprl}.
For finite $U$, the Lieb-Wu solution \cite{liebwu}
shows there is a charge gap for arbitrarily small
$U$ and thus one expects Mott-Hubbard order for
any positive $U$, but with a magnitude vanishing
as $U \rightarrow 0$.
This is consistent with our numerical results
(Fig. \ref{fig:su2su3}) for the behavior of the
singlet insertion/deletion correlator.

In contrast, for the simplest $SU(3)$ generalization of
the Hubbard model, a finite critical
interaction strength, $U_c$, should be required
to induce a metal-insulator transition \cite{charles}.  
Our numerical results for this model are consistent
with the absence of ODLRO for $SU(3)$ singlet insertion
for $U < U_c \sim  2 t$, and the
presence of  ODLRO for $SU(3)$ singlet insertion for
larger values of $U$.

We can also make contact between our singlet insertion
operator and the bosonization picture of the one dimensional
Mott-Hubbard transition;
in the large $U$ limit of the
$SU(2)$ Hubbard model, the singlet insertion operator at site $j$
is effectively acting like a constant times $(-1)^{\sum_{k<j} n_k}$,
where $n_k$ is the
number of fermions at site $k$.  In
the Luttinger liquid approach to the Hubbard model, 
$(-1)^{\sum_{k<j} n_k}$ maps to $\cos(\Theta_{\rho,J})(ja)$.
The two point, equal time correlation function of
this operator decays asymptotically like $x^{-1}$ for 
free particles, as does the singlet-insertion singlet-deletion
correlator, and, in the Luttinger liquid approach
to the Hubbard model, $\cos(\Theta_{\rho,J})$ 
takes on an expectation value $\propto (-1)^j$ for
any finite $U$, with the magnitude vanishing as $U\rightarrow0$
via a Kosterlitz-Thouless transition.  This further
supports the singlet insertion operator acting as an order parameter
for the one dimensional Mott-Hubbard transition and
we feel confident that our operator is an effective
probe of the Mott-Hubbard transition in one dimension.

What does this picture imply for
the Mott-Hubbard transition in higher dimensions?
First, note that the singlet insertion operator
reveals the presence of a broken $Z_2$, {\em spatial\/} symmetry.
Such a broken symmetry doubles the
unit cell and might be expected to lead,
for a half-filled system, to a filled band
and a gap to
all excitations; however, if the broken symmetry involves
only the {\em charge\/} degrees of freedom,
then the result is an anti-ferromagnetic
insulator rather than a band insulator, which would have
gapped spin degrees of freedom.  This behavior is reminiscent of
the formation a spin density wave (SDW)
insulator, however, in that case
the ``order parameter'' transfers real electrons
across the Fermi surface and therefore couples to both
spin and charge degrees of freedom. Further, the low lying spin
degrees of freedom of the SDW are Goldstone bosons resulting from the
broken $SU(2)$ symmetry and therefore the spin-density
wave picture can not apply in one spatial dimension
or in two dimensions at any finite temperature \cite{mermin}.  The
breaking of a discrete symmetry, such as $Z_2$,
is allowed in one dimension at $T=0$ and 
causes the Mott-Hubbard transition.  Such a breaking
is also allowed {\em at finite temperature in two or
more dimensions\/}.  This suggests that higher dimensional
Mott-Hubbard transitions in pure Hubbard models
may well be finite temperature
transitions at which a $Z_2$ symmetry breaks
in the charge sector, leaving a low energy effective theory
containing only magnetic degrees of freedom.  This contrasts
sharply with the conventional wisdom on the Mott-Hubbard transition
which follows the line of argument that, if the only
order parameter for the Mott-Hubbard transition is
the conductance, then no true, finite temperature distinction exists 
between metal and insulator \cite{SDW}.
This argument fails if one is prepared to admit the
possibility of the breaking of a $Z_2$ symmetry 
only in the charge sector.
Based on our one dimensional result that the
Mott-Hubbard transition does admit a pure charge order parameter
which breaks a discrete symmetry,
we feel that this proposal should be seriously considered.

Interestingly, two organic conductors, (TMTTF)$_2$SbF$_6$
and (TMTTF)$_2$ReO$_4$ \cite{pouget}
exhibit abrupt changes in their
charge properties consistent with finite
temperature metal-insulator transitions, and, moreover,
{\em these transitions are unaccompanied by any detectable
magnetic, structural or charge density wave transitions\/}.
Existing explanations for the transition between
metallic and insulating behavior in similar materials \cite{emery}
are based on a finite temperature crossover that can not
account for such sharp ``transitions'', thus it
appears very natural to interpret this behavior as
finite temperature Mott-Hubbard transitions 
of the type we are proposing.

To test whether the exotic sort of ordering we are proposing
actually occurs requires a suitable probe for Mott-Hubbard
order beyond one dimension.  In this case, the singlet insertion
operator we have used is no longer appropriate, however, it
can be generalized into a useful probe.  Consider the
two dimensional case of a Hubbard model of size
$L_x \times L_y$ : 
\begin{eqnarray}
H_{\rm 2D ~Hubb} & = & 
 \sum_{ x,y,\sigma,\nu}
t_{\parallel}  c_{\sigma,x+\nu,y}^{\dagger} c_{\sigma,x,y} + 
t_{\perp} 
c_{\sigma,x,y+\nu}^{\dagger} c_{\sigma,x,y}  \\
\nonumber & + & 
U \sum_{x,y} n_{\uparrow,x,y} n_{\downarrow,x,y}
\end{eqnarray}
where we identify $x = L_x +1$ with $x = 1$, with $L_x$ an even integer,
and $y = L_y+1$ with $y = 1$.
Now, we 
define a multi-singlet insertion operator
which inserts $L_y$ singlet pairs, all at fixed $x$, and
study the dependence on $L_x$ of the expectation value of
this operator acting at $x=0$ and the conjugate removal
operator acting at $x=\frac{L_x}{2}$.
If the system has completely anisotropic hopping 
($t_{\perp} = 0$) so
that each of the $L_y$ chains is independent then,
at zero temperature and for positive $U$,
this expectation value
behaves like the $L_y$th power of the one dimensional
singlet insertion correlator.  For large $L_x$ and odd $L_y$,
this depends on $L_x$ only as $(-1)^{\frac{L_x}{2}}$,
reflecting the underlying one-dimensional Mott-Hubbard
order.  Now imagine that we turn on some finite
interchain hopping, $t_{\perp}$, with $U \gg t_{\parallel}$
and $U \gg t_{\perp}$.    We can use the multi-singlet insertion
to see if the Mott-Hubbard order persists at finite $t_{\perp}$.

\begin{figure}
\narrowtext
\centerline{ \epsfxsize = 2.5in
\epsffile{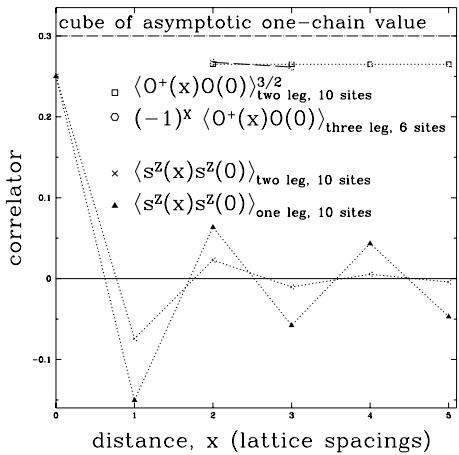}}
\caption{Results of multi-singlet insertion/deletion correlator
for two and three leg spin ladders (PBC, and
$J_{\rm out~of~chain} = 2 J_{\rm in~chain}$). Three leg ladder
shows alternation, and both two and three leg ladders
appear to have ODLRO for $O$ with the magnitude of
$\langle O^{\dagger} O \rangle$ only slightly reduced 
from that of uncoupled chains. Contrast this with
the dramatic effects of the interchain coupling on
$\langle s^z s^z \rangle$,
which, for example, switches from algebraic to exponential decay for the two leg
ladder.} 
\label{fig:threechains}
\end{figure}

As a first step in this direction, we considered 
multi-singlet insertion for coupled spin chains;
the low energy theory for the two dimensional
Hubbard model at sufficiently large $U$ should be that
of the Heisenberg  model:
\begin{equation}
H =  \sum_{x,y}J_{\parallel}
\vec{S}(x,y) \vec{S}(x+1,y)+
J_{\perp} 
\vec{S}(x,y) \vec{S}(x,y+1)
\end{equation}
with $J_{\parallel} \sim \frac{t_{\parallel}^2}{4U}$
and $J_{\perp} \sim \frac{t_{\perp}^2}{4U}$. 
The interchain
spin couplings are relevant operators \cite{meandy}
and could destroy the singlet insertion order;
however, recall that our order parameter acted as a
$c$-number in the spin sector. In this case,
the usual arguments about perturbed conformal field theories 
imply that its behavior should show no infrared divergences
caused by the relevant interchain couplings \cite{anotherarg}. 
Therefore the behavior of our multi-singlet
insertion operator should be be qualitatively 
the same as the isolated chains case: $\propto (-1)^{\frac{L_x L_y}{2}}$.  
If so, the two dimensional
Heisenberg model, and, by inference,
presumably the two dimensional Hubbard model in the
Mott insulating phase, possess hidden, broken $Z_2$
symmetries.  We have
numerically studied the behavior
of the multi-singlet insertion correlator for
two and three leg spin chains with the results depicted in
Fig. \ref{fig:threechains}.
\vspace*{-0.6cm}
As can be seen from the figure, the relevance of the interchain coupling 
does not lead to any strong change in behavior in the
multi-singlet insertion operator. For the three leg case
a broken $Z_2$ appears to persist; 
this suggests that such a discrete symmetry breaking may
be present in the two dimensional Hubbard model and
underly its Mott-Hubbard transition.
In the future, we hope that
it may be possible to use the multi-singlet insertion to resolve the question
of the existence of a finite temperature
$Z_2$ breaking in this model
and we are currently exploring this possibility.

In summary, we have proposed and given evidence in
support of the proposal that an operator which inserts a
pair of singly occupied sites in a singlet spin configuration
into a one dimensional Hubbard model acts as an
order parameter for the Mott-Hubbard transition.  The transition
occurs in this language because of the breaking of a spatial $Z_2$
symmetry in the charge sector of the model, doubling the charge unit
cell and causing an insulator.  A natural conjecture based on this
picture is that a similar Mott-Hubbard transition can occur in
higher dimensions,
involving the breaking of a charge sector
$Z_2$ symmetry, possibly at finite temperature.
In support of this, we have given some numerical evidence for 
a broken $Z_2$ for the two dimensional Heisenberg model
and identified two experimental systems whose behavior is
suggestive of the proposed, new kind of ordering.

S. P. S. gratefully acknowledges support from DOE grant
DE-FG02-90ER40542.


\end{multicols}

%
%

%
%


\begin{references}
\bibitem{SDW} Naturally, if the metal-insulator transition
is tied to some more conventional transition,
then the order parameter for
that transition can be used.  See also Ref. \protect{\cite{georgesfootnote}} 
where a first order, finite temperature metal-insulator
transition was found.
\bibitem{anotherarg} In fact, the robustness of the order to relevant interchain
spin couplings provides another argument in support of
the proposal that our singlet insertion/deletion is
probing a charge sector order.
\end{references}
\end{document}